# ML-based Secure Low-Power Communication in Adversarial Contexts

1st Guanqun Song
song.2107@osu.edu
*The Ohio State University*

2nd Ting Zhu
zhu.3445@osu.edu
*The Ohio State University*

*Abstract*—As wireless network technology becomes more and more popular, the mutual interference between various signals has become more and more severe and common. Therefore, there is often a situation in which the transmission of its own signal is interfered with by occupying the channel. Especially in the confrontation environment, Jamming has caused great harm to the security of information transmission. So I propose ML-based secure ultra-low power communication, which is an approach to use machine learning to predict future wireless traffic by capturing patterns of past wireless traffic to ensure ultra-low-power transmission of signals via backscatters. In order to be more suitable for the adversarial environment, we use backscatter to achieve ultra-low power signal transmission, and use frequency-hopping technology to achieve successful confrontation with Jamming information. In the end we achieved a prediction success rate of 96.19%.

## I. Motivation

With the development of science and technology, machine learning techniques have been applied to various fields such as medicine, business, security and so on [1–8], especially in the direction of computer vision and natural language processing have made remarkable achievements[9–15]. Therefore, in the field of wireless signals, the predictive ability of machine learning can also be used to prevent possible interference threats. At the same time, in the confrontation environment, the transmission of information should be real-time and accurate, and the ultra-low power consumption should be used to ensure normal communication for a long time. Therefore, we propose ML-based secure ultra-low power communication applicable under various protocols [16–25] to ensure ultra-low power transmission while effectively avoiding the attack of interfering signals and achieve normal communication against the environment [27–30]. The system architecture is shown in the Fig.1.

## II. Description of the Questions

Nowadays, various wireless signals are ubiquitous, and they follow specific protocols under normal conditions and peacefully coexist in an environment. In the ISM public frequency band, frequency is a very precious resource. As shown in the Fig.2, the 2.4GHz frequency band includes WiFi, Bluetooth and ZigBee, as well as various signals. In this way, it is necessary to avoid other signal interference, and prevent the signal from being unable to be sent normally because the channel is occupied.

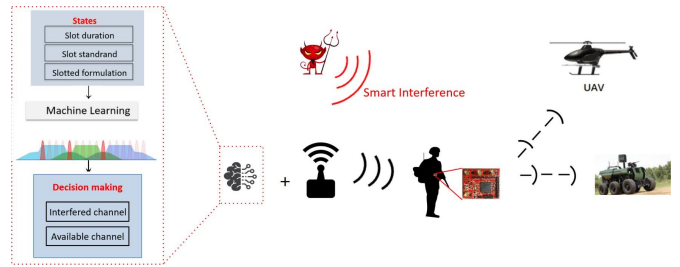

Fig. 1. System Architecture

However, in adversarial contexts, the attacker will use various ways to interfere with the signal the sender wants to transmit. Once the channel is occupied by the jamming signal, the transmitted signal will be lost, which will have very serious consequences in an adversarial environment. At the same time, the power consumption of electrical equipment should also be considered. An intermediate medium with ultra-low power consumption that is easy to carry around should be used to transmit the signal, and the encryption and security of the signal can be guaranteed.

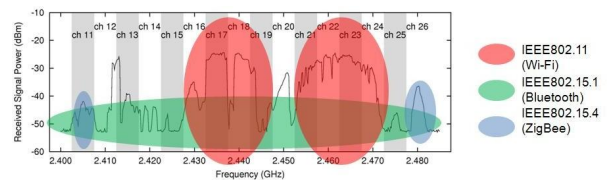

Fig. 2. Wireless Signal Channels

## III. Related Work

Adversarial wireless communication is a complete combat network system with network center and informatization as the core, which requires accurate and efficient realization of intelligence, coordination and situational connectivity in adversarial scenarios. Communication confrontation can be divided into communication interference and anti-jamming according to the nature of the function. Communication interference mainly delays the enemy's information transmission, compresses the enemy's communication distance and range, and even destroys the enemy's information transmission link. With the increasingly complex electromagnetic environment of the battlefield, the modern informationized battlefield requires that the electronic warfare communication system needs to have strong anti-interference ability, which can conceal the combat attempt and improve the suddenness and concealment of military operations.

The best interference is one that completely suppresses the enemy's wireless communications. Its basic principle is that the interference signal can cover the communication signal in the multi-dimensional space such as time domain, frequency domain, power domain and spatial domain, and the interference signal and the waveform of the communication signal are related to realize the suppression in the multi-dimensional space. Communication jamming technology can be divided into targeting jamming and blocking jamming, and jamming jamming includes broadband/partial frequency band jamming jamming, comb jamming jamming, sweeping collision jamming jamming and hopping collision jamming jamming[31].

In previous work, some people have investigated jamming of sorts. Hossein Pirayesh and Huacheng Zeng [32] investigated and sorted out the specific performance of jamming and anti-jamming in the common wireless network environment in life, and explained the reasons and proposed improvement methods from the physical layer. At the same time, in order to effectively use the performance of machine learning to solve the jamming problem, Han et al. [33] proposed to use the anti-jamming method to ensure that Zigbee can successfully send signals in multiple Zigbee and Wifi environments. In order to realize the construction of data sets for many different OFDM signals and the construction of jamming data sets, I have learned many mathematical formulas for simulation from this book [34], and will use the many different waveforms recorded in the book to train to obtain a more general meaningful machine learning model. In addition, Xin et al. [35– 42] used backscatter to modulate the environmental signal. In my paper, I found that the frequency-shifting technology of backscatter can be used to realize the channel hopping of the signal, and then the frequency-hopping can avoid the jamming attack according to the prediction result of machine learning, and finally achieve success. Communication, and because backscatter does not need to use batteries, it can use radio frequency signals for spontaneous charging and data transmission, which can achieve the goal of ultra-low power transmission. My proposed idea is built on OFDM, so the scope of application of this method can be easily further trained to use LTE or 5G, etc., so it will mainly focus on various jamming situations of OFDM [43] for more in-depth research in the future. In addition, this solution can be used for UAVs and self-driving cars to form a communication network to achieve a high degree of decentralization and parallelism [44–48], eventually creating a new effective communication architecture [50–60].

## IV. PROPOSED SOLUTION

Regarding the proposed solution, I want to achieve the final experimental goal through the following steps:

- Propose machine learning strategy to learn jamming attack pattern.
- Predict attacker's optimal action in both time domain and frequency domain.
- Identify the best channel and time slots for backscattering signals.
- Dynamically change the backscattering channel to mitigate the jamming attacks.
- Combine the frequency-shifting of backscatter and the prediction ability of machine learning to achieve accurate prediction and ultra-low power transmission.

## V. DESCRIPTION

It is feasible to use machine learning to predict whether the signal has interference, because the signal has completely different waveforms at two different times. By learning regular waveforms and abnormal waveforms through CNN, the probability prediction of interference channels in a specific time period can be performed, so as to obtain channels that can be used for signal transmission.

At the same time, when the channel is occupied, we can perform frequency hopping on the signal through the prediction result. As shown in Fig.3, there is noise interference in fb.17 fb.20, because the frequency hopping technology can be used to avoid the interference channel and continue communication. Because the backscatter has the characteristic of frequency shifting, the backscatter is used to reflect the signal into the idle channel so that the signal can communicate successfully.

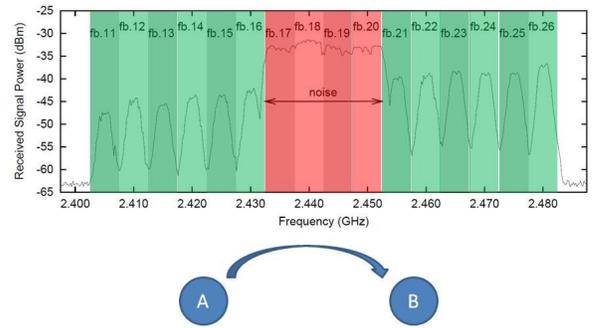

Fig. 3. Frequency Hopping

## VI. DESIGN

In order to build a training machine learning model dataset, I simulate various conventional OFDM signals to make a standard signal dataset, and expand the interference dataset by simulating various interference signals, and divide it into 80% as the training set according to the experimental needs, 20% as a test set. The specific types of interference signals are as the following:

## A. Single Tone Jamming

Single Tone Jamming usually refers to audio interference with only one frequency point. The mathematical model is as follows:

$$j(t) = \sqrt{2J} \cos\left(2\pi f_j t + \theta_j\right) \quad (1)$$

where J is the average power of the tone, $f_j$ is its frequency, and $\theta_j$ is its phase offset from the target signal.

The time-domain graph and power spectrum of the single-tone interference are shown as Fig.4 and Fig.5. Because the single-tone interference only acts on a small frequency band with a very narrow bandwidth, it cannot effectively interfere with the frequency hopping system. However, because of the characteristics of concentrated interference power, it has great advantages in the interference of narrowband signals.

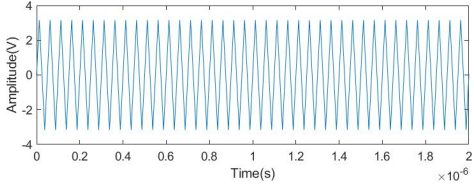

Fig. 4. Single-tone Interference Signal Waveform

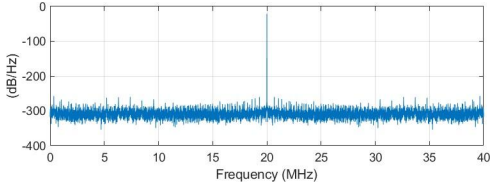

Fig. 5. Single-tone Interference Signal Spectrum

## B. multi-tone interference signal

Multi-tone interference is multiple single-tone interference, and its expression is as follows:

$$J_M(t) = \sum_{i=1}^{M} \sqrt{J_i} \exp\left[j\left(2\pi f_{J_i} t + \varphi_i\right)\right] \quad (2)$$

where $f_{J_i}$ represents the frequency of the $i^{th}$ interference tone; $\varphi_i$ represents the power of the $i^{th}$ interference tone; $\varphi_i$ is the random phase uniformly distributed in [0~ $2\pi$]. In the interference, the $f_{J_i}$ needs to be selected within the frequency band, otherwise there will be no interference. The time-domain graph and power spectrum of the multi-tone interference are shown as Fig.6 and Fig.7

## C. Linear sweep interference signal

Linear swept interference is also known as chirp interference. Its instantaneous frequency changes linearly with time, it can be regarded as a single tone signal at a certain time point, and it exhibits the characteristics of broadband signal and dynamic scanning at a certain time period. The time

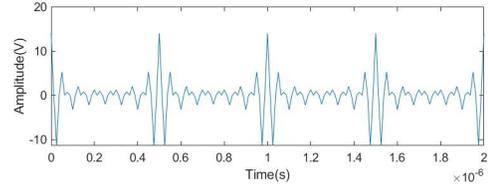

Fig. 6. Multi-tone Interference signal waveform

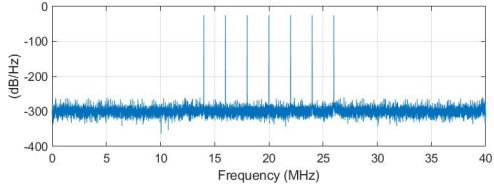

Fig. 7. Multiple-tone Interference Signal Spectrum

domain expression of the linearly swept interference signal is as follows:

$$j(t) = A \exp\left[j\left(2\pi f_0 t + \pi k t^2 + \varphi\right)\right] \quad 0 \leq t \leq T \quad (3)$$

where A is the amplitude of the interfering signal, $f_0$ is the initial frequency of the interfering signal, k is the frequency modulation coefficient of the interfering signal, $\varphi$ is the initial phase of the signal, and T is the duration. Among them, the amplitude of the interference signal and the frequency modulation coefficient are important parameters that affect the performance of the frequency sweep interference. The time-domain graph is shown as Fig.8.

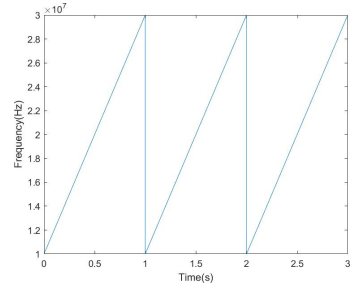

Fig. 8. Linear sweep signal waveform

## D. Sawtooth Sweep Interference Signal

The swept-frequency jamming signal is a broadband jamming signal with time division in both the frequency domain and the time domain. The principle is to use a narrow-band interference signal with a relatively narrow bandwidth to scan a wider interference frequency band in a frequency sweep period. We set the narrowband interference signal as the RF noise interference:

$$X(t) = U_n(t) \cos\left(\omega_j t + \varphi(t)\right) \quad (4)$$

where $U_n(t)$ is the signal envelope, $\omega_j$ is carrier frequency, $\varphi(t)$ is a uniform distribution of $[0, 2\pi)$. Then the mathematical model of frequency sweep interference is as follows:

$$J(t) = U_j X(t) \cos(2\pi f_j t + \theta_t + \varphi) \quad (5)$$

where $U_j$ is the amplitude of the swept-frequency interference interference signal, $\varphi$ represents the initial phase of the interfering signal. The instantaneous frequency of the interference signal is as follows:

$$f(t) = 2\pi f_j + \frac{d\theta(t)}{dt} = 2\pi f_j + F(t) \quad (6)$$

$F(t)$ is a frequency sweep function, and the frequency sweep function is a continuous signal. The frequency sweep function generally uses the sawtooth wave function, which satisfies the following formula:

$$F(t) = 2\pi kt, \, nT_f \le t \le (n+1)T_f, \, n = 0, 1, 2 \cdots \quad (7)$$

where k is the frequency sweep slope, and $T_f$ is the frequency sweep period. The time-domain graph and power spectrum of the sawtooth swept interference signal are shown as Fig.9 and Fig.10

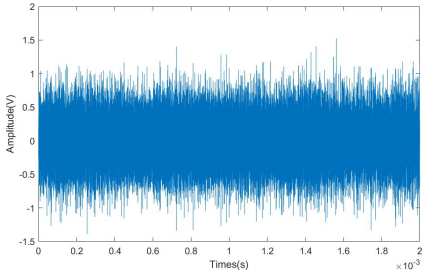

Fig. 9. Saw-tooth sweep signal waveform

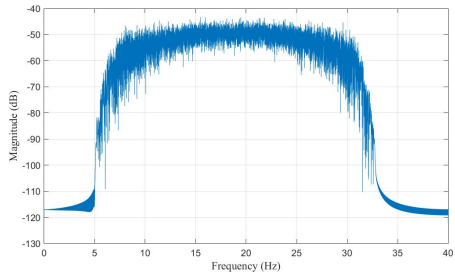

Fig. 10. Sawtooth swept signal spectrum

In the experiment, broadband blocking interference is used to interfere with the communication signal, so the signal will be affected for a long time in the whole communication process. The formula is as follows:

$$\Delta f_j > 5\Delta f_r, f_s \in [f_j - \Delta f_j/2, f_j + \Delta f_j/2] \quad (8)$$

where $\Delta f_j$ is the interference bandwidth, $\Delta f_r$ is the single-channel bandwidth, $f_s$ is the communication center frequency, and $f_j$ is the interference center frequency. I changed slot duration, slot standard and slotted formulation by changing parameters and make the pos and neg dataset as Fig.12 and Fig.13.

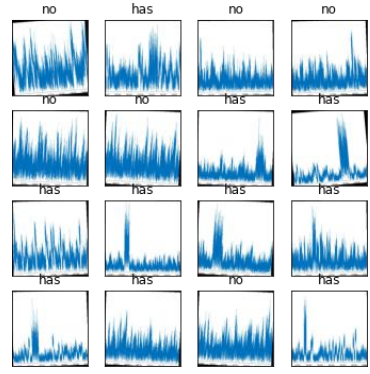

Fig. 11. images in training test

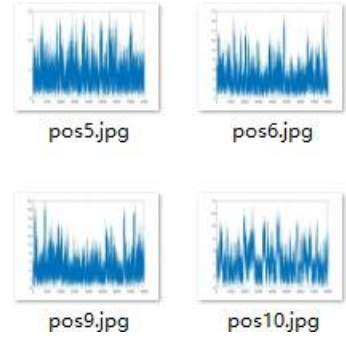

Fig. 12. Conventional Signal

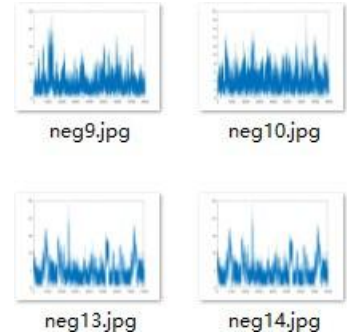

Fig. 13. Interfered Signal

## VII. EVALUATION

In the data preprocessing stage, we want our model to identify the images correctly irrespective of the size and positioning of an object in the image. I set up transforms.Resize((350)) to resize the images so the shortest side has a length of 350 pixels. The other side is scaled to maintain the aspect ratio of the image. And use transforms.CenterCrop(300) to crop the center of the image so it is a 300 by 300 pixels square image. Then using transforms.ToTensor() converts my image into numbers. Finally, Use transforms.Normalize to subtract the mean from each value and then divides by the standard deviation to preprocessing 4398 images of all simulations. And I label the generated image data by manual labeling. The training set data labeling results are shown as Fig.14.

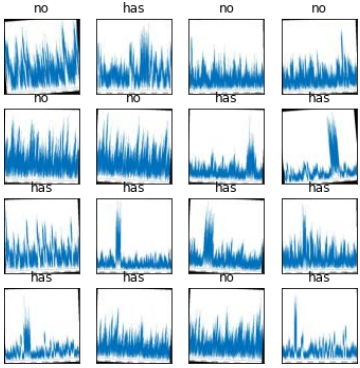

Fig. 14. Images and Labels in Training Test

For the convolutional neural network, the definition of neural network is shown in the Fig.15. I use two convolutional layers, one pooling layer, 3 fully connected layers, using SGD as the optimizer for training. Because of the dataset, when the epoch reaches 50, the loss will basically not decrease. The minimum loss of the training set is equal to 0.038082. The model performance test is performed on the test data set, and the Test Loss obtained for 1002 pictures is 0.028825 as TableI.

```
Net(
  (conv1): Conv2d(3, 16, kernel_size=(7, 7), stride=(1, 1))
  (pool): MaxPool2d(kernel_size=2, stride=2, padding=0, dilation=1, ceil_mode=False)
  (conv2): Conv2d(16, 32, kernel_size=(7, 7), stride=(1, 1))
  (dropout): Dropout(p=0.2, inplace=False)
  (fc1): Linear(in_features=156800, out_features=256, bias=True)
  (fc2): Linear(in_features=256, out_features=84, bias=True)
  (fc3): Linear(in_features=84, out_features=1, bias=True)
)
```

Fig. 15. CNN Net

TABLE I
MODEL LOSS

|      | Train Loss min | Test Loss |
|------|----------------|-----------|
| Loss | 0.038082       | 0.028825  |

We use commonly used machine learning performance evaluation metrics to evaluate the model, mainly focusing on Precision, Recall, and F-Measure. The performance of the model in the test set is shown in Fig.16.

Finally, the manual validation set is used to judge the usability of the model. We used a total of 122 images for testing, and only 2 images were not correctly predicted, and the accuracy rate was 98.361%. Finally, the manual validation set is used to judge the usability of the model. We used a total

```
classification_report
              precision    recall  f1-score   support

           0       0.97      1.00      0.99       501
           1       1.00      0.97      0.98       501

    accuracy                           0.99      1002
   macro avg       0.99      0.99      0.99      1002
weighted avg       0.99      0.99      0.99      1002
```

Fig. 16. Performance Evaluation

of 122 images for testing, and only 2 images were not correctly predicted, and the accuracy rate was 98.361%. The reason for not being correctly predicted is still that the waveforms are very similar and the signal-to-noise ratio is very high. , causing the model to judge incorrectly. Two of the examples are as follows Fig.17 and Fig.18.

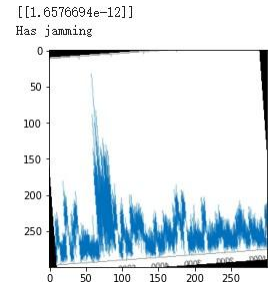

Fig. 17. Has Jamming

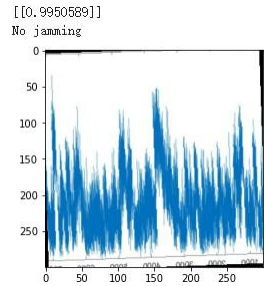

Fig. 18. No Jamming

VIII. ISSUES YOU ENCOUNTERED

At the beginning of the data set construction, the prediction rate reached 95% in 10 epochs, because the data set was too regular. Therefore, modifying different slot durations and different occurrence times as parameters to expand the data set enables the machine learning model to adapt to more and more complex situations.

Later, the trained machine learning model was tested for prediction, and it was found that some interference signals were still judged as channel occupancy when the strength of the interference signal was weak, but this did not conform to the basis for manual judgment. Therefore, the error analysis is carried out, and the picture with the wrong prediction is subdivided and classified, and the strong interference and weak interference are classified in detail, and a more applicable and accurate model is obtained.

Finally, I want to use the trained model to get a machine learning model that fits the real signal by training it with real signal data. But the bandwidth of the oscilloscope is 200MHz, which means the minimum time length is 5 nanoseconds. This means that even if the signal is actually simulated, our device cannot detect the signal, and better equipment is needed to detect and output images for training, so it is difficult to train the model from the simulated signal to the real signal. But the analog signal can still prove the availability of the system.

## IX. Conclusion

In this paper, we propose ML-based Secure Low-Power Communication in Adversarial Contexts, a cross-domain structure that leverages machine learning to make predictions and use the results to evade malicious interference signals. In our experiments, the prediction results generated by the training model based on the simulation dataset can reach higher than 96%, which proves that the structural design is feasible, and can use machine learning to successfully predict the signal, providing a new channel detection method , and introduce an effective strategy for predicting signal interference.

## X. Future Work

In the future, I need to complete the wireless communication system first, complete the scattering of the signal through backscatter and the correct demodulation of the signal to ensure the correct transmission of the signal. At the same time, different from the single backscatter in this experiment, the coordinated scattering of multiple backscatters is considered, and the receiver needs to be able to deploy a machine learning model to split and decode the mixed scattered signals to achieve further machine learning and wireless communication. Combined to form a complete machine learning-based ultra-low power wireless communication transmission system in adversial contexts.